\documentstyle[preprint,aps]{revtex}

\begin{document}

\thispagestyle{empty}

\title{Vanishing condensates and anomalously light \\
Goldstone modes in medium}

\author{Thomas D. Cohen}

\address{Department of Physics and Astronomy, University of Maryland,
College Park, MD 20742, USA}

\author{Wojciech Broniowski}

\address{H. Niewodnicza\'nski Institute of Nuclear Physics,
         PL-31342 Cracow, POLAND}

\maketitle

\begin{abstract}

We show that in a uniform medium, the vanishing of a particular
condensate along with spontaneously broken symmetry   imply the
existence of an anomalously light pseudo-Goldstone mode.  The
consequences for a vanishing chiral condensate in nuclear matter
are discussed.

\end{abstract}

\newpage


Recently, Ericson \cite{ericson:ref} observed that if the  chiral
condensate, $\langle \overline{q} q \rangle$, goes  to zero in nuclear
matter at some density, one does not know {\it a priori} that chiral
restoration occurred.
The possibility exists that $\langle \overline{q} q \rangle$ can vanish without
chiral restoration.
In this letter, we show from very
general arguments that if Ericson's scenario of a vanishing condensate
without symmetry restoration is realized, then the pion's mass must be
anomalously light in a sense which
will be precisely defined.

The letter is organized as follows:
We begin by introducing the problem in the
very general context of the ground state (subject
to a spatially uniform constraint) of
an arbitrary  quantum field theory with a nearly conserved current.
Next we note the connection of this general problem to the problem of
nuclear matter with
a vanishing chiral condensate.  We proceed to demonstrate for the
general case that a vanishing condensate without symmetry restoration must
imply
an anomalously light pseudo-Goldstone boson.   Finally, we discuss the
implications of this result in the context of  nuclear matter and
simple pion condensation models and show why our results do not apply
for the kaon condensation problem.  We also briefly address the role of spatial
correlations in  $\overline{q} q$ in this problem.   Before proceeding  we
wish to remind the reader that all of the results in this paper apply
only to spatially uniform systems such as infinite nuclear matter

Consider a field theory
with a Noether current, $J_{\mu}$,
which is nearly conserved.  That is, the divergence of the
current can be written as
\begin{equation}
\partial^{\mu} J_{\mu} = \lambda D,
\label{broken}
\end{equation}
where $D$ is a local operator and $\lambda$ is a parameter in the
Lagrangian which can be made arbitrarily small. The theory has an exact
symmetry in the limit $\lambda \rightarrow 0$.  The symmetry is
spontaneously broken if $\lim_{\lambda \rightarrow  0^+} \langle [Q,
D]_{\rm ET} \rangle \neq 0$, where the subscript ET indicates equal
time, $Q = \int {\rm d}^3x J_0$, and the brackets indicate a
ground-state expectation value. The quantity $\lim_{\lambda
\rightarrow  0^+} \langle [Q, D]_{\rm ET} \rangle$ is called the {\em
condensate}, and is often thought of as the order parameter associated
with the symmetry breaking \cite{condensate:end}.  As is well known,
spontaneous symmetry breaking (in the absence of a Higgs mechanism)
implies the existence of massless Goldstone bosons
\cite{Goldstone:ref}.  Moreover,  as shown by Gell-Mann, Oakes and
Renner (GMOR) \cite{GMOR:ref},  if the explicit breaking parameter,
$\lambda$, is nonzero then  the square of the mass of the {\em
pseudo-Goldstone} boson is proportional  to $\lambda$.

Implicit in the usual derivation of the GMOR relation is the fact that
the condensate is negative semi-definite.  Thus, a condensate of zero
is special.  Of course, a vanishing condensate obviously occurs when
the symmetry is not spontaneously broken.  In the present {\em Letter},
however,  we consider a different possibility:  suppose the condensate
vanishes, {\em i.e.}
\begin{equation}
\lim_{\lambda \rightarrow  0} \langle [Q, D]_{\rm ET}
\rangle = 0 \; ,
\label{condition}
\end{equation}
but that nevertheless the symmetry remains spontaneously broken.  We
will demonstrate here that this implies that the corresponding
pseudo-Goldstone bosons must be anomalously light, {\it i.e.} the energy
of the pseudo-Goldstone modes increases with $\lambda$ more slowly than
$\lambda^{1/2}$.  This means that
\begin{equation}
\lim_{\lambda \rightarrow 0+} \frac{E_{\rm PGB}^2}{\lambda} = 0
\end{equation}
(where $E_{\rm PGB}$ is the energy of the pseudo-Goldstone boson),
rather than the usual case, in which this limit gives a nonzero value.
In fact, many years ago it was suggested that the chiral condensate of
QCD vanishes in the vacuum despite the existence of spontaneous chiral
symmetry breaking \cite{qcdcond:ref}.  In this scenario the square
of the pion
mass is proportional to the square of the  current quark mass rather
than  to the current quark mass itself.  It is believed that this
scenario is quite implausible in the context of QCD since it violates
a  naturalness condition --- it depends  on the condensate  vanishing
``accidently''\cite{GL:ref}.

The principal new result in this {\em Letter} is that the fact that a
vanishing condensate in the presence of spontaneous symmetry breaking
implies that the pseudo-Goldstone modes are anomalously light holds
more generally than for the vacuum.  In particular, the result can be
generalized  to any spatially uniform ground state of the theory
(medium) subject to a constraint (such as a spatially uniform
 density of some conserved
current) so long as the operator which specifies the constraint
commutes with $Q$.
This generalized condition has an important possible application in
nuclear physics.  The underlying dynamics is, of course, QCD, and the
current of interest is the axial vector current.  The condensate is
$\langle \overline{q} q \rangle$, and the role of $\lambda$ is played
by the current quark mass $m_q$.   Because this result holds only for
spatially uniform systems, we  consider the problem of  infinite
 nuclear matter at nonzero baryon density; since the constraint must
commute with the chiral charges (which carry isospin)
 we restrict our attention to isospin symmetric nuclear matter.

There is general agreement that as the baryon density of infinite  nuclear
matter increases, the chiral condensate decreases
\cite{ericson:ref,CFG:ref,DL:ref,LKW:ref,CSDZ:ref,ME:ref,CM:ref,BM:ref,MB:ref}.
It is possible that at a sufficiently high density the condensate
vanishes.  This may be a signature of the
chiral restoration phase transition.  However, as recently observed in
a model by Ericson \cite{ericson:ref}, the fact the condensate vanishes
does not necessarily imply that the symmetry is restored --- the
expectation value of other chiral symmetry breaking operators may be
nonzero.  The relevance of the present general result in this context
is clear:  if Ericson's scenario is correct, and the chiral condensate
vanishes without chiral restoration,  then the pseudo-Goldstone modes,
{\it i.e.} the pions \cite{pions:end}, must be anomalously light.  Such
anomalously light pions are what one expects in the vicinity of an {\em
S-wave pion condensation} phase transition.  The existence of
anomalously light pions in this situation is a nontrivial dynamical
consequence deduced from general principles.  Of course, without
solving QCD at finite densities one does not know whether Ericson's
scenario occurs in the real world. However, this scenario is realized
in simple models  of S-wave pion condensation based on  mean field
treatment of the nonlinear $\sigma$ model. By construction, these
models always have spontaneous symmetry breaking.  At some density the
condensate vanishes.  Simultaneously, the pion mass becomes anomalously
light --- indeed, it goes to zero and pion condensation occurs. Note that
in the context of nuclear matter it is not ``unnatural'' for the
condensate to vanish without chiral restoration, since the vanishing
condensate is associated with a phase transition.

Now we come to the core of this {\em Letter}.  Let us consider  the
general case, in which the state of interest is the ground state of the
theory subject to a spatially uniform constraint. In our initial
discussion we will ignore issues associated with renormalization.  As
we will argue later, the renormalization effects do  not affect our
conclusions.  The constraint is  imposed by insisting that the
expectation value of some local constraint operator, $C(x)$, in the
state of interest is equal to some fixed value independent of $x$.
Here we consider the case where the constraint operator commutes with
the charge associated with the Noether current $J$, that is
\begin{equation}
[Q, C(x) ]_{\rm ET} \, =
\, [\int {\rm d}^3 y \, J_0(y), C(x) ]_{\rm ET} \, = \,
0 \label{cond}
\end{equation}
Let us denote the {\em ground state} subject to the constraint
by $|C \rangle$ (the vacuum is simply  the special case for which $C=0$).
Since, our constraint is spatially uniform,  the constraint commutes
with the three-momentum.  We define
\begin{equation}
G_B^C(\lambda) \equiv \, i \, \langle C | [Q, B(0)]_{\rm ET} |
 C \rangle_{\lambda} \; ,
\label{CB}
\end{equation}
where B is an arbitrary local operator  independent of $\lambda$,
and the subscript $\lambda$ indicates
a fixed value of $\lambda$ and a fixed constraint.
We assume that the symmetry is spontaneously
broken, which implies that there exists some operator $B$ for which
$\lim_{\lambda \rightarrow  0^+} G_B^C (\lambda) \ne 0$.

The key issue in the analysis is the insertion of a complete
set of intermediate states into Eq.~(\ref{CB}).  Since the medium
is spatially uniform, these states can be labeled by their three-momentum
relative to the state $|C\rangle$,
and an additional label $j$.   Now let us denote the energy of the
the state relative to the state $|C\rangle$ as  $E_j(\vec{p})$.  Note
$E_j$ and $\vec{p}$ are {\it not} the absolute energy and momentum
of the state, but the difference of the energy and momentum of the
state from the ground state.  For example, if the state
is a single quasiparticle state,  $E_j(p)$ and $\vec{p}$ are
the natural variables to describe the energy and momentum of the quasiparticle.
Since the total energy and total momentum of the state labeled $j$ and of
the ground state (subject to constraints) are both four-vectors,
their difference is also a four-vector; hence $E_j$ and $\vec{p}$ form
a four-vector and $E_j^2 - p^2$ is a Lorentz scalar.
Unity can be decomposed as
\begin{equation}
{1} = \sum_j  \int \frac{{\rm d}^3p }{(2 \pi)^3 2|E_j(p)|}
   |j, {\vec{p}} \rangle \langle j, {\vec{p}}| \, ,
\label{unity}
\end{equation}
where the factor $(2 |E_j(\vec{p})|)^{-1}$ is included to ensure the Lorentz
invariance of the measure, and the states  have the standard covariant
normalization $ \langle j, \vec{k}| j, \vec{p} \rangle = 2
|E_j(p)| \delta(\vec{p} - \vec{k})$.  Note the absolute value
signs on the energy.  Of course, if $|C\rangle$ is the vacuum, then
$E_j$ must be  positive.  Of course, for $C \ne 0$ there are, in general,
states lower in energy than $|C \rangle$.  As we shall see, however,
given the assumption that our constraint commutes with the charge,
intermediate states with $E_j < 0$ will not contribute to
$G_B^C(\lambda)$.

Inserting Eq.~(\ref{unity}) into Eq.~(\ref{CB}) yields
\begin{equation}
\label{CB2}
G_B^C(\lambda)  = - \sum_j \, E_j^{-1} \, {\rm Im} \left (
\langle C |J_0(0) | j , \vec{p}=0 \rangle_{\lambda}
\langle j ,\vec{p}=0|B(0)|C\rangle_{\lambda} \right ) \; ,
\end{equation}
where $E_j$ implicitly refers to the state with $\vec{p}=0$.
Note that at this step we have dropped the absolute value sign.
The reason is the condition in  Eq.~(\ref{cond}) that the
charge in question commutes with the constraint, plus the fact
that $|C\rangle$ is the minimum energy state subject
to the constraint.  Thus, the only states coupled to $|C\rangle$
by $Q$ are states with an energy greater than or equal
to the energy of $|C\rangle$.

One can use Eq.~(\ref{broken})
to relate   matrix elements of $J_0$ in Eq.~(\ref{CB2})
to matrix elements of $D$:

\begin{equation}
\langle j ,\vec{p}=0| J_0(0)| C\rangle_{\lambda} =
- i \, \frac{\lambda}{E_j} \, \langle j ,\vec{p}=0|
D(0)|C\rangle_{\lambda} \; . \label{relate}
\end{equation}
Inserting this into Eq.~(\ref{CB2})    gives
\begin{eqnarray}
\label{CB3}
G_B^C(\lambda) = - \sum_j \, \frac{\lambda}{ E_j^2} \,
{\rm Re} \left ( \langle C |D(0) | j , \vec{p}=0 \rangle_{\lambda}
\langle j ,\vec{p}=0|B(0)| C \rangle_{\lambda} \right ) \; .
\end{eqnarray}
An important special case is where the operator $B$ is itself $D$.
In this case Eq.~(\ref{CB3}) becomes
\begin{equation}
G_D^C(\lambda) = \, - \, \sum_j \, \frac{\lambda}{ E_j^2} \, |
\langle C |D(0) | j , \vec{p}=0 \rangle_{\lambda} |^2 \; .
\label{CD}
\end{equation}
The limit of $G_D^C (\lambda)$ as $\lambda \rightarrow 0^+$ is, by
definition, the condensate.  To proceed we need to make the rather
innocuous assumption that the matrix elements
\mbox{$\langle C |D(0) | j , \vec{p}=0\rangle_{\lambda}$}
and
$\langle C |B(0) | j , \vec{p}=0\rangle_{\lambda}$
do not diverge as $\lambda \rightarrow 0^+$.
Given this assumption, the condensate can be written as
\begin{equation}
\lim_{\lambda \rightarrow 0^+} G_D^C(\lambda) =
- \sum_j  | \langle C |D(0) | j , \vec{p}=0 \rangle_{\lambda=0} |^2
\lim_{\lambda \rightarrow 0^+} \, \frac{\lambda}{ E_j^2} \; .
\label{CD2}
\end{equation}
Note that every term on the right-hand side of Eq.~(\ref{CD2}) is
negative semi-definite.  Thus, we observe that {\em the condensate is
negative semi-definite}. This is a rather general corollary of our
derivation, which extends the vacuum result to the constrained state
$|C\rangle$.  It is worth noting that this result depends in an essential way
on the fact that our constraint commutes with the charges so that the only
states connected have an energy greater than or equal to that of $|C\rangle$.

Furthermore, if the condensate is nonzero, there must exist at least
one term in the sum which is nonzero.  Given, the assumption that
$\langle C |D(0) | j , \vec{p}=0 \rangle_{\lambda=0}$ is finite, one
concludes there must be at least one term in the summation for which
$\lim_{\lambda \rightarrow 0^+}  \frac{\lambda}{ E_j^2} \neq 0$.  Such
a mode is a pseudo-Goldstone mode.  As expected, the square of the energy
of this state is proportional to the symmetry breaking parameter,
$\lambda$.  Equation (\ref{CD2}) may be viewed as an in-medium
generalization of the GMOR relation.

Now consider the situation in which the condensate is zero.  Since each
term in Eq.~(\ref{CD2}) is negative semi-definite, this implies that
the limit of each  term  as ${\lambda \rightarrow 0^+}$   must be
zero.  One solution is that the symmetry has been restored.
 Let us suppose, however, that the symmetry remains spontaneously
broken.  This means that there exists some operator $B$ for which
$\lim_ {\lambda \rightarrow 0^+} G_B^C(\lambda) \ne 0$, which implies
that at least one term in Eq.~(\ref{CB3}) is nonzero.  For both of
these situations to be simultaneously true, there must exist some mode
$j$ for which
\begin{equation}
\label{1}
\lim_ {\lambda \rightarrow 0^+} \frac{\lambda}{ E_j^2} \,
\cos(\phi_{BD}^j) | \langle C|D(0) | j , \vec{p}=0
\rangle_{\lambda} | \, | \langle j ,\vec{p}=0|B(0)|C \rangle_{\lambda} |\,
= \, a_j \, \neq 0 \; ,
\end{equation}
\begin{equation}
\lim_ {\lambda \rightarrow 0^+}   \frac{\lambda}{E_j^2} \,
| \langle C |D(0) | j , \vec{p}=0 \rangle_{\lambda} |^2  \,
= \, 0 \; ,
\label{2}
\end{equation}
where $\cos(\phi_{BD}^j) $ is the relative phase between
the matrix elements of $D$ and $B$.
Since by hypothesis $\langle j ,\vec{p}=0|B(0)|C\rangle_{\lambda = 0} $
is not divergent,  Eq.~(\ref{1}) can be  satisfied only if
\begin{equation}
\lim_ {\lambda \rightarrow 0^+}  \frac{\lambda}{ E_j^2} \,
| \langle C |D(0) | j , \vec{p}=0 \rangle_{\lambda} |
 = \frac{a_j}{\cos(\phi_{BD}^j)  \, | \langle j ,\vec{p}=0|B(0)| C
\rangle_{\lambda=0} | }  \; .
\label{3}
\end{equation}
Equations~(\ref{2}) and (\ref{3}) together imply
$\lim_ {\lambda \rightarrow 0^+}  |
\langle C |D(0) | j , \vec{p}=0 \rangle_{\lambda} |  = 0$,
which combined with Eq.~(\ref{3}) implies that
$\lim_ {\lambda \rightarrow 0^+}  \frac{\lambda}{ E_j^2}  \rightarrow \infty$.
This completes the demonstration that spontaneous symmetry breaking with a
vanishing condensate implies the existence of an anomalously
light mode --- {\em i.e.} a mode for which
\begin{equation}
\lim_{\lambda \rightarrow 0^+} \frac{E_j^2}{\lambda} = 0 \; .
\end{equation}

As mentioned previously, the preceding discussion ignored
renormalization.  It should be apparent, however, that renormalization
effects will not alter the results.  The operators $J_{\mu}$, $D$ and
$B$ should be taken to be the appropriate renormalized operators.  The
only issue which needs to be addressed is the fact the operator product
in Eq.~(\ref{cond}) needs to be renormalized.  Since we are ultimately
treating these operator products via summation over intermediate
states, as in Eq.~(\ref{CB2}), the renormalization of the composite
operator is equivalent to the inclusion  of ultraviolet subtraction
constants in the summation.  The key point in the present analysis,
however, is that in the limit $\lambda \rightarrow 0$, the only
intermediate states contributing to the sum are the pseudo-Goldstone modes.
This is clear from Eq.~(\ref{CB2}-\ref{relate}) where the overall
factor of  $\lambda$ causes all terms to vanish in this limit except
terms associated with modes whose mass goes to zero as $\lambda
\rightarrow 0$.  Thus, in this limit, there are no contributions from
high mass states and hence no  ultraviolet divergences.  Since the
result derived here is  formally valid only in the limit $\lambda
\rightarrow 0$, it is clear that it is not destroyed by
renormalization.

Now, let us proceed to some applications of our general result.  In the
context of $SU(2) \times SU(2)$ chiral symmetry in spatially uniform
isoscalar nuclear
matter, we have three partially conserved axial vector currents,
$J^a_5$, the parameter $\lambda$ is the average up and down current
quark mass, $m_q$, the condensate is $\lim_{m_q \to 0^+} \langle
\overline{q} q \rangle$, and the (isospin degenerate) Goldstone modes
are the pions: $\pi^0$, $\pi^+$ and $\pi^-$.  Thus, if $\langle
\overline{q} q \rangle =0$ and chiral symmetry is not restored, the
pion becomes anomalously light.  As noted above, this is what one
expects in the vicinity of an s-wave pion condensation transition.

We can also draw conclusions for isospin asymmetric nuclear matter, but only
for the properties of the {\em neutral} pion, $\pi^0$.  This is because
the constraint operator, which in this case is the third isospin
component of the vector charge, $J^{3}$, commutes only with the third
isospin component of the axial vector charge, $Q_5^3$, and does not
commute with $Q_5^1$ and $Q_5^2$.  Hence we are only allowed to rotate
about the third isospin direction, and the relevant symmetry is the
$U(1) \times U(1)$ subgroup of the full chiral $SU(2) \times SU(2)$.
Thus, the assumptions of our theorem are met for the case of the
neutral pion, which becomes anomalously light if the condensate
vanishes and the chiral $U(1) \times U(1)$ remains spontaneously
broken.

Of course, it is an open question for infinite uniform nuclear mattter
in QCD whether $\lim_{\lambda \to 0^+}\langle \overline{q} q \rangle$
ever goes to zero without chiral restoration.
However, this scenario  can be realized  in simple models such as
mean field treatments of the nonlinear $\sigma$ model.  Such
models have been used to explore the possibility of both
$\pi$ and $K$ condensation \cite{MPW,PW,KN,BKRT,BLRT}.
Here, let us consider an $SU(2) \times SU(2)$
version of the model.  The principal building block of the theory
is the chiral field
\begin{equation}
U = \exp \left ( i \frac{\pi \cdot \tau}{f_{\pi}} \right ) \; ,
\end{equation}
where $\pi$ is the pion field and $\tau$ represents the Pauli matrices.
The commutator of $U$ with the axial vector charge is given by
 \begin{equation}
 i \, [Q^5_a , U] \, = \, i \, f_{\pi} \,  \partial_{\pi_a} U \; .
\end{equation}
In these models, $U$ is taken to be the only chirally active degree of
freedom; all other degrees of freedom (such as baryons) commute with
the axial vector charge. Thus, for these models the condensate, which
we denote $\langle \overline{q} q \rangle$  in analogy to the
equivalent quantity in QCD, is given by
\begin{equation}
\langle \overline{q} q \rangle \, = \, \frac{ \langle f_{\pi}^2
\partial^2_{\pi_a} {\cal L}^{\chi s b} \rangle }{ m_q} \; ,
\end{equation}
where ${\cal L}^{\chi s b}$ is the chiral symmetry breaking part of the
sigma model Lagrangian.  It is therefore apparent that a vanishing
condensate implies $\langle \partial^2_{\pi_a} {\cal L}^{\chi s b}
\rangle = 0$, which means that the pions have zero excitation energy.
 A zero pion energy
is consistent with our result since it is anomalously light.  In a
simple mean field model \cite{simple:model} with heavy nondynamical
nucleons ({\it i.e.} ignoring all terms second order in the baryon
density) this occurs at the density of $m_{\pi}^2 f_{\pi}^2 /
\sigma_{\pi N}$,  where $ \sigma_{\pi N}$ is the $\pi$-N sigma term.
In the context of such models, this is taken to be the point at which
s-wave pion condensation occurs.  As stressed by Yabu, Myhrer and
Kubodera\cite{simple:model},  however, this result is highly model
dependent and can easily be altered by the inclusion of effects second
order in the density.

We have yet to address an extremely important issue.  Our general result
was derived in the limit that $\lambda \rightarrow 0^+$.  In applications to
chiral symmetry in nuclear physics the parameter is some combination of
quark masses and these are small but not zero.   Clearly, if the masses
are small enough
the physics will be essentially unchanged  from the zero mass limit.
The important question is ``How small is small enough?''  Going back to the
derivation the answer is clear.  The result applies
qualitatively, so long as the
 state $|C \rangle$ for
a finite quark mass is essentially the same as the state
 in the massless quark limit (in the sense that the various matrix elements of
 $|C \rangle$ which enter the derivation are nearly the same as for the
 massless quark case).
This restriction is very important.  While we have discussed the case
of s-wave pion condensation, we have not addressed the more interesting
case of s-wave kaon condensation which is generally believed to be far
more probable.   Our general result, unfortunately, tells us nothing about
this problem.  The difficulty is that in the context of our derivation
the strange quark mass cannot be regarded as small.  The problem of
interest is the transition at some critical density from ordinary
nuclear matter (with zero net strangeness) to a condensed phase.
However, in the limit $m_s, m_u , m_d \rightarrow 0$
 ordinary nuclear matter is never the ground state of the theory
 subject to a constraint of fixed baryon density;  SU(3) symmetric
 matter, with the strange quark density equal to 1/3 of the baryon
density is the ground state.  Thus, the ordinary nuclear matter state
for physical quark masses is qualitatively changed as one goes to the
chiral limit and  our general result tells us nothing.  One might
suppose that the problem can be circumvented by
using the density of up and down quarks (rather than the baryon
density) as the constraint.  While this does prevent the ground state from
changing qualitatively as one goes to the chiral
limit, it still does not allow
the application of our general result since the axial currents containing
strange quarks (which are connected to the modes of interest for kaons)
now do not commute with the constraint which violates a necessary condition
for our theorem. The case of non-commuting constraints will be reported
elsewhere \cite{tdcwb:to:be}.

We have demonstrated a very general result and shown its implications
to descriptions of nuclear matter.  We have not yet addressed one important
issue, the role of correlations in $\langle \overline{q} q \rangle$.  The
essential physical point stressed by Ericson\cite{ericson:ref} is that
the chiral condensate is an average quantity.  In a reasonable
description of nuclear matter there are very strong spatial
correlations in the local value of $\langle \overline{q} q (x)\rangle$.
 In essence, one expects that near (and inside) nucleons the condensate
is very different from its vacuum value while between nucleons
the value of $\langle \overline{q} q(x) \rangle$ may not be much
altered from the vacuum.  Given this picture, it is reasonable to ask
whether this average
quantity should play any special role.  It apparently represents the
average of two very different kinds of physics and one might imagine that it
could
go to zero or even change signs without anything special (such as chiral
restoration) occurring.  For this to happen, it seems that all  that must occur
is that the positive contributions in the vicinity of nucleons has to become
larger than the negative contributions between nucleons.

The formal derivation outlined in this letter does not specifically address
the question of correlations.  However, it does show that intuition based
on the existence of strong correlations can be misleading.  The chiral
condensate --- the averaged quantity --- is special.   For example, from
very general considerations, one sees that it is negative semi-definite (to
leading
order in $m_q$).  A vanishing chiral condensate, despite its average
nature, has strong implications for the dynamics: implying either chiral
restoration or anomalously light pions.

To summarize, we have shown that for the ground state of a system
subject to a constraint which commutes a nearly conserved current, a
vanishing condensate implies either that the symmetry is not
spontaneously broken or that the pseudo-Goldstone modes are anomalously
light.  In the context of nuclear physics, this implies that if the
chiral condensate vanishes for isoscalar nuclear matter at some
density, then this density is at or near the critical density for
either an S-wave  pion condensation phase transition or a chiral
restoration phase transition.

We acknowledge useful discussions with G. E. Brown, M. K. Banerjee and
H. Forkel.  This work has been supported by the NSF--Polish Academy of
Science grant \#INT-9313988, NSF PYI grant \#PHY-9058487, DOE grant
\#DE-FG02-40762, and partly by the Polish State Committee for Scientific
Research grants 2.0204.91.01 and 2.0091.91.01,
by the Maria Sk\l{}odowska-Curie grant PAA/NSF-94-158,
and by the Alexander von Humboldt Foundation.

\end{document}